\documentclass{sigchi}
\setcopyright{acmcopyright}
\usepackage{balance}       % to better equalize the last page
\usepackage{graphics}      % for EPS, load graphicx instead 
\usepackage[T1]{fontenc}   % for umlauts and other diaeresis
\usepackage{txfonts}
\usepackage{mathptmx}
\usepackage[pdflang={en-US},pdftex]{hyperref}
\usepackage{color}
\usepackage{booktabs}
\usepackage{textcomp}

\usepackage{microtype}        % Improved Tracking and Kerning
\usepackage{ccicons}          % Cite your images correctly!
\usepackage{todonotes}

\def\plaintitle{A Hybrid Approach for Data Analytics for Internet of Things}

\def\emptyauthor{}
\def\plainkeywords{Internet of Things; Cloud; Data Analytics; Fog Computing; Edge Computing; Distributed Data Analytics}

\makeatletter
\def\url@leostyle{%
  \@ifundefined{selectfont}{
    \def\UrlFont{\sf}
  }{
    \def\UrlFont{\small\bf\ttfamily}
  }}
\makeatother
\def\pprw{8.5in}
\def\pprh{11in}

\definecolor{linkColor}{RGB}{6,125,233}
\hypersetup{%
  pdftitle={\plaintitle},
  pdfauthor={\emptyauthor},
  pdfkeywords={\plainkeywords},
  pdfdisplaydoctitle=true, % For Accessibility
  bookmarksnumbered,
  pdfstartview={FitH},
  colorlinks,
  citecolor=black,
  filecolor=black,
  linkcolor=black,
  urlcolor=linkColor,
  breaklinks=true,
  hypertexnames=false
}

\begin{document}

\title{\plaintitle}

\numberofauthors{3}
\author{%
  \alignauthor{Badraddin Alturki\\
    \affaddr{University of Leicester}\\
    \affaddr{Leicester\\ United Kingdom}\\
    \email{baba1@leicester.ac.uk}}\\
  \alignauthor{Stephan Reiff-Marganiec\\
    \affaddr{University of Leicester}\\
    \affaddr{Leicester\\ United Kingdom}\\
    \email{srm13@leicester.ac.uk	}}\\
  \alignauthor{Charith Perera\\
    \affaddr{Newcastle University}\\
    \affaddr{Newcastle upon Tyne\\ United Kingdom}\\
    \email{charith.perera@ncl.ac.uk}}\\
}

\maketitle

\begin{abstract}
The vision of the Internet of Things is to allow currently unconnected physical objects to be connected to the internet.  There will be an extremely large number of internet connected devices that will be much more than the number of human being in the world all producing data. These data will be collected and delivered to the cloud for processing, especially with a view of finding meaningful information to then take action. However, ideally the data needs to be analysed locally to increase privacy, give quick responses to people and to reduce use of network and storage resources. To tackle these problems, distributed data analytics can be proposed to collect and analyse the data either in the edge or fog devices. In this paper, we explore a hybrid approach which means that both in-network level and cloud level processing should work together to build effective IoT data analytics in order to overcome their respective weaknesses and use their specific strengths. Specifically, we collected raw data locally and extracted features by applying data fusion techniques on the data on resource constrained devices to reduce the data and then send the extracted features to the cloud for processing. We evaluated the accuracy and data consumption over network and thus show that it is feasible to increase privacy and maintain accuracy while reducing data communication demands. 
\end{abstract}

\category{D.2.11.}{Software Engineering}{Software Architectures --- Data Abstraction} \category{H.3.4.}{Information Systems}{Systems and Software --- Distributed Systems} 


\section{Introduction}
The Internet of Things (IoT) has become one of the most active areas in computer science and beyond, both for researchers and companies and it is interpreted by communities in a variety of ways. The cluster of European research projects defined the IoT as allowing "people and things to be connected Anytime, Anyplace, with Anything and Anyone, ideally using Any network and Any service \cite{Vermesan2009}". This is important for applications to monitor, track, and communicate (amongst others) with things for various purposes remotely. Sensors embedded in many everyday physical objects around us play a key role in IoT. These embedded sensors will include vast sensing capabilities \cite{7004800} and can then send the data through the network to decision points -- typically the cloud. After collecting the data, there is a need to analyse them to gain insights to help, automate and speed up decision making\cite{preden2015data}. In a data driven economy, the data and insights can be considered as the main goods \cite{ng2014engineering}. According to \cite{sundmaeker2010vision} the number of devices connected to the internet will be more than 50 billion devices in the very near future. However, the greater awareness promised by so many smart things will produce an ever greater volume of data at increasing rates of delivery. 

Big data is not a new term in computer science \cite{zaslavsky2013sensing}, it has been created by big technological companies like Yahoo, Microsoft and Google. Big data as researched has three key characteristics: volume, variety and velocity \cite{zaslavsky2013sensing}. As result of improvements in electronics, the cost of equipment has decreased dramatically and sensors have become more affordable and are already embedded into many electronic devices. A great number of data has been generated by these sensors and companies have started storing it -- in fact the predominant business model at the moment seems to be around storing and owning data for possible later analytics: most fitness trackers or smart watches will send their recorded data to the cloud of the maker. Part of this desire to store comes from the value of data, part of the fact that data analytics is a notable challenge to which solutions are still being explored. However, the IoT moves the game to an entirely new level by increasing the scale of deployed devices dramatically -- posing new challenges to gathering, processing, transporting, storing and analysing data.

There are five steps (Collection, Collation, Evaluation, Decide, and Act) in the so called IoT monitoring cycle \cite{wang2015city}. Considering the overall IoT systems, we always have devices at the edge as well as in the network and the cloud in a central position. Data processing can be at the in-network (so edge and devices in the network) and the cloud level \cite{abu2013data,wang2015city} -- and these levels play a role in the various stages. 

Most of the research and existing work in the field of big data focuses on cloud computing because of the offered power in terms of processing and storage. The common way to process the data is to send all data to the cloud and return results after analysis. In addition to the significant power available, processing in the cloud also means that as complete a collection of data is available to analysis as can be obtained. However, processing all streaming raw data in the cloud negatively effects several aspects, such as increased network traffic, latency (to get actions back to the user), energy consumption and privacy. As the IoT grows the need to tackle these issues grows.

We suggest that in the longer term there is an opportunity to move the computation as much as possible off the cloud to the fog or edge device side. This means that data analytics should be handled in the device or fog before sending the data to cloud (possibly going as far as avoiding the cloud altogether for processing of operational data). The cloud would still have a role in longer term backup and also in helping to compute models to guide analytics, in fact there is no doubt that the cloud has an important role in enabling the IoT since it provides high power processing and storage \cite{DBLP:journals/corr/abs-1301-0159}. A key problem to be tackled is to understand how accuracy of analytic results is effected if computations and decisions on raw data are made elsewhere in the processing chain and infrastructure.

So, in this paper we propose a hybrid approach that moves some processing off the cloud and allows us to study the savings in data transfer and changes to accuracy. In the proposed work, we are fusing and filtering data close to the source and then send meaningful higher level data rather than raw data to the cloud. As less details are being transmitted some privacy protection (not every little move is known, only the general picture) is already taking place -- however further work in studying the privacy angle needs to be undertaken.

This data fusion technique will directly influence the collation and evaluation steps and hence the crucial question arising is: how can we fuse sensors data locally without harming the accuracy of the overall decision? A secondary question is considering the feasibility of distributing the processing considering that many network and edge devices have less processing power.

The novel contributions of this paper are:
\begin{itemize}
\item We propose a hybrid approach, which moves the computation as much possible to fog/ edge side of the network. 

\item We extensively evaluate the approach using the WISDM dataset \cite{kwapisz2011activity} and five of the most popular data analytics techniques.

\item We explore the feasibility of applying these data aggregation techniques via resource constrained device, particularly a Raspberry Pi 3 Model B.
\end{itemize}

The rest of this paper is organised  as follows: section 2 describes the solution space while section 3 explores our solution. We then evaluate, consider related work and draw conclusions.

 \section{Solution Space: Data Aggregation}
CERP-IoT \cite{sundmaeker2010vision} provides the following characteristics of the IoT: Autonomous, Intelligence, connectivity, sensing, energy, dynamism, interoperability, privacy and security. The characteristics of data in the IoT are heterogeneity, redundancy, dynamism and variety. Considering that "data fusion and mining present an efficient way to manipulate, integrate, manage and preserve mass data collected from various things" \cite{yan2015trustworthy}. Processing IoT data means to add value to the raw data by extracting important aspects and creating meaningful information -- an essential element of the IoT \cite{xhafa2014semantics}, \cite{andrade_gedik_turaga_2014} identifies five steps to follow when processing IoT data, namely data collection, data pre-processing, transformation of data, mining and evaluation. 
In this paper we are specifically interested in data fusion, which fits into the area of data pre-processing and transformation and allows to reduce the volume data but increase its value. Data fusion is referred to by other 'synonyms' such as information fusion, decision fusion, data combination, multi-sensor data fusion, sensor fusion and data aggregation. While there is no general agreement on these terms, there are some differences that can be observed: in some cases data fusion is applied on raw sensor data while information fusion is used to determine analysed data, meaning that the latter has a higher semantic grade than data fusion\cite{castanedo2013review}. Similarly, data fusion techniques are used to integrate data from a variety of sources to produce more meaningful and effective inferences and associations, whereas data aggregation can be considered as subcomponent of data fusion which summarises the sensor data to remove data redundancy \cite{abdelgawad2012data}. The most common definitions by researchers are as follows:
\begin{itemize}
\item  data fusion is defined by the Joint Directors of Laboratories (JDL) workshop \cite{white1991data} as "a multi-level process dealing with the association, correlation, combination of data and information from single and multiple sources to achieve refined position, identify estimates and complete and timely assessments of situations, threats and their significance." 
\item  Hall and Llinas \cite{hall1997introduction} say that "data fusion techniques combine data from multiple sensors and related information from associated databases to achieve improved accuracy and more specific inferences than could be achieved by the use of a single sensor alone."
\end{itemize}

\begin{figure*}[htb]
  \centering
 % \vspace{-0.83cm}
  \includegraphics[width=16cm]{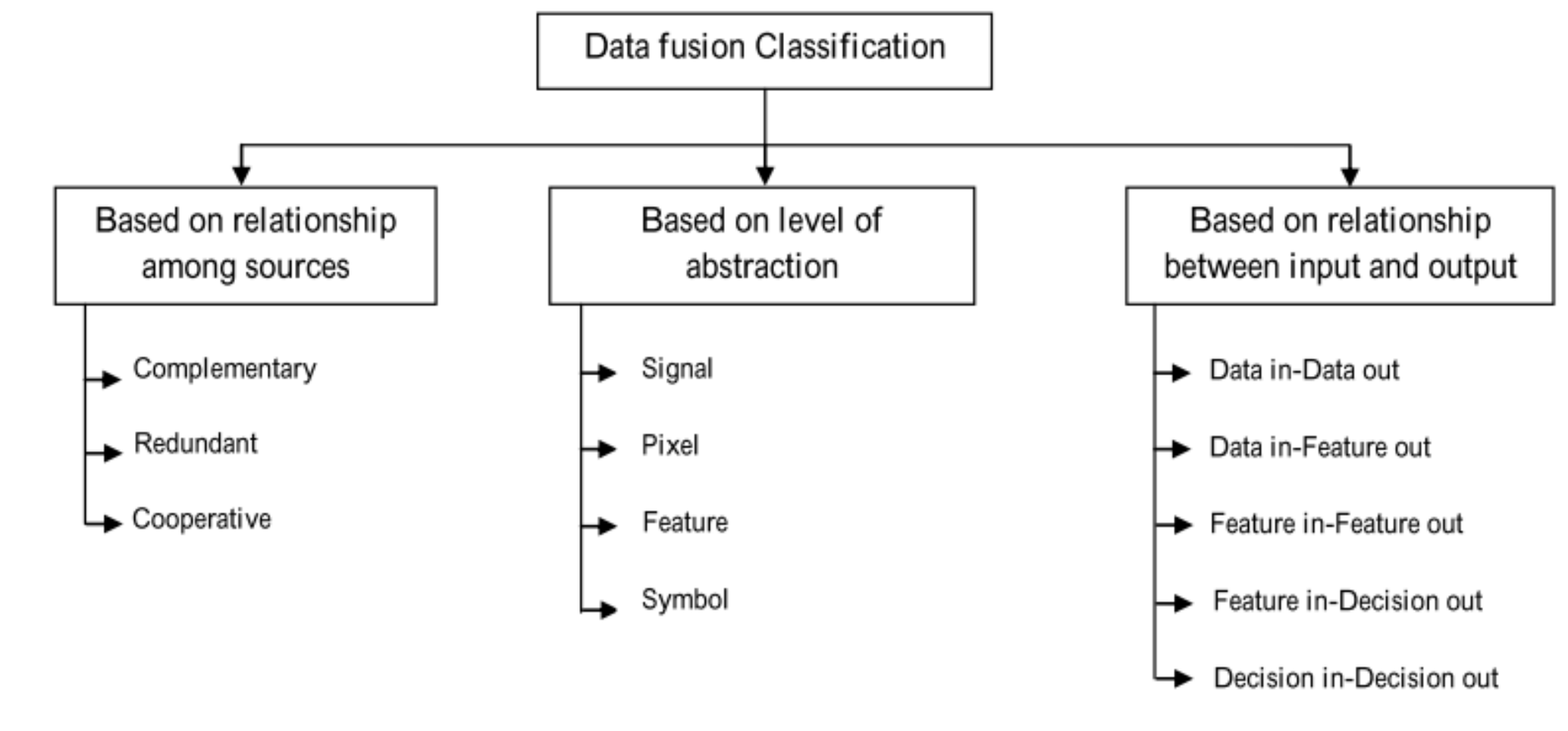}
 % \vspace{-0.21cm}	
  \caption{Data Fusion Classification}
 \label{Figure:Data Fusion Classification}	
 %\vspace{-0.23cm}	
 \end{figure*}

Data fusion can be classified depending on a variety of attributes as shown in figure \ref{Figure:Data Fusion Classification} \cite{chhabra2015data}. These attributes are discussed in detail in \cite{abdelgawad2012data} and generally capture the idea that there are different dimensions such as the abstraction level or the relation between the data items from one or multiple sensors.

Dasarathy's data fusion classification system formalises the attributes just discussed and can be considered as one of the most common approaches \cite{dasarathy1997sensor}. Dasarathy's classification focuses on details of input and output based on the abstraction level. The classification contains five classes as follows \cite{castanedo2013review}): 

\begin{description}
\item [Data In-Data Out (DAI-DAO)] is the primary method of data fusion in the classification model. It processes the raw data that are collected directly from sensors  resulting in more accurate data. In addition, image and signal processing algorithms can be used at this stage.
\item [Data In-Feature Out (DAI-FEO)] processes the raw data to produce features which can depict a structure about the environment.
\item [Feature In-Feature Out (FEI-FEO)] processes a collection of features to get more effective feature results.
\item [Feature In-Decision out (FEI-DEO)] processes the features to acquire a collection of decisions.
\item [Decision In-Decision Out (DEI-DEO)] processes the decisions to extract more efficient decisions.
\end{description}

Features are defined as the single measurements that are used to create the training model. In other words, they are the columns of data that are created for the training set \cite{harrington2012machine}. In addition, data fusion can provide the required knowledge that is essential in a decision-making process, therefore, the amount of the available knowledge / data can effect the final decision at any stage. Many techniques use symbolic information and the data fusion process to determine the uncertainties and restrictions that are part of / effect the decision-making process \cite{castanedo2013review}. In other word, the decision can be captured depending on the knowledge of the events that are collected from variety sources by fusing them.

Informally, our working definition of data fusion can be that it aggregates and integrates all sensor data to allow obtaining accurate and meaningful data while eliminating unneeded and useless data. 

Understanding what data fusion can achieve, one also needs to consider the architectural aspect of where data fusion is applied. Options include a centralised, decentralised or distributed architectures as follows \cite{castanedo2013review}:

\begin{itemize}
\item Centralized architecture: all the collected data from sensors will be sent to the cloud for processing which means that everything is held in one single server. It is known that the cloud is capable to process very large amounts of data effectively. However, in real time scenarios data consumption over the network will be high, which will make the cloud not sufficient for effective fusion of the data. This architecture is also very problematic if the data consists of images such as earth observation imagery. The reason is that there will be more delays in terms of data arrival time and this will impact badly on the output of data. Additionally, privacy will be one of the main issues because this architecture receives all the raw data without applying any reduction or aggregation previously. Finally, energy consumption has been important in IoT because transferring raw data all time from devices using any network such 3G and WiFi will consume significant amounts of energy.

\item Decentralized architecture: there are several nodes in the network and each of them has their specific computation capabilities, so there is no single server like centralised system. Every node applies data aggregation autonomously on its local data and data received from peers. One of the major limitations of this architecture is the high communication cost between peers. In this case, if we increase the number of nodes, then there might be a lack of scalability.

\item  Distributed architecture: sensor readings are processed at the source level before applying data aggregation in a specific node that is capable of data fusion. This can overcome various issues of the centralised architecture and can reduce communication costs over the decentralized architecture.

\item Hierarchical architecture: The data fusion step is performed at a variety of levels in the hierarchy and it can be considered as a combination of both distributed and decentralised architecture.
\end{itemize}

It is true that it is not possible to say that one of these architecture is the best, as it often depends on specific requirements and technology. Both decentralised and distributed architectures are quite similar to each other in many ways. However, they differ in terms of the place for pre-processing the data. In decentralised architectures the whole data aggregation happens in every node which produces comprehensive output. Whereas, in distributed architectures the raw data is firstly pre-processed at source to extract features, and then these features are fused. The main advantages of the distributed architecture over the centralised one are reducing the processing and communication costs because it pre-processes the data in a distributed manner before fusing data \cite{castanedo2013review}.

It is generally accepted that increasing accuracy and reducing energy usage are major aspects of data fusion \cite{chhabra2015data}, so any architecture that is presented needs to consider these aspects. While accuracy is self explanatory, reducing energy is more difficult as the energy used is a combination of costs for storage, transport and processing with transport being very expensive on wireless transmissions technologies.

As there are obvious trade-offs between the different architectures it seems desirable to formulate solutions which combine the different ideas in ways that reduce the disadvantages and benefit from the advantages of each. Our method presented below attempts to achieve this.

  \section{Proposed Solution: Adaptive Data Aggregation}

  \subsection{Overview}

   \begin{figure*}[htb]
  \centering
 % \vspace{-0.83cm}
  \includegraphics[width=18cm]{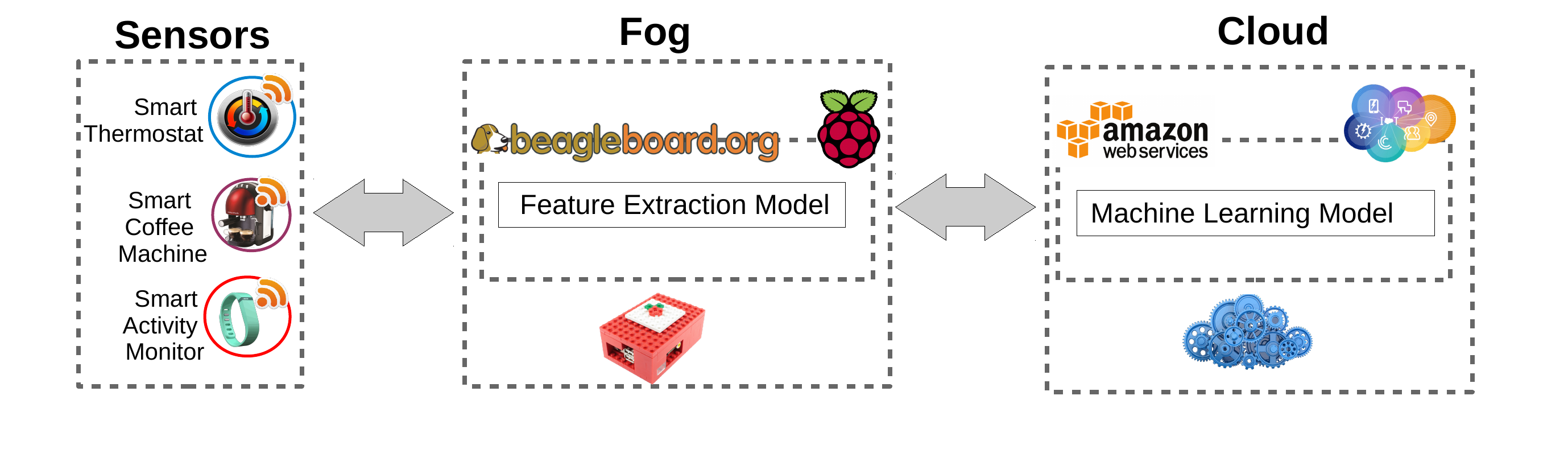}
 % \vspace{-0.21cm}	
  \caption{Distributed Processing in Internet of Things}
 \label{Figure:Distributed Processing in Internet of Things}	
 %\vspace{-0.23cm}	
 \end{figure*}

We propose a hybrid approach that moves the computation as much as possible from the cloud to the fog/edge level. The overview of this approach is demonstrated in Figure \ref{Figure:Distributed Processing in Internet of Things}. We begin with applying data fusion techniques on sensors data to minimise the number of data points and extract features in IoT devices. Then, we extract features from this data and send it to edge/fog node. This step is important because it is widely accepted that raw time-series data cannot be efficiently analysed by ordinary algorithms for classification. After that, the features will be sent to the cloud for training purposes and creating inferences.

  \subsection{Architecture}
  
The architecture of our proposed solution is divided into two main parts:  First, the cloud level has the responsibility of data training and creating inferences. Second, the in-network level aggregates the sensor data to reduce data transmission cost over network. The aim is to save energy, reduce decision times and to increase privacy by analysing and processing data locally while maintaining accuracy as much as possible, reduce data transmission cost over network and save energy by analysing and processing data locally. 

The communications between the nodes or peers can be undertaken in different ways as is typical (such as WiFi, 3G and any other solutions). Figure 2 shows the architecture of the system. The distributed processing architecture contains three types of node including IoT Devices, fog, and the cloud as follows:

\begin{itemize} 
\item A \textbf{sensor node} is at the lowest level of the system and is typically embedded in physical objects. Sensor nodes are small and cheap in terms of price to make the process of deploying sensors to objects easy and inexpensive. It senses real-world inputs such as motion detection, temperature and so on. These sensor nodes are connected to Fog nodes via wireless or wired communication. 

\item A \textbf{fog node} resides next to the sensors or along the communication path to the cloud and collects sensors data (or data received from a 'downstream' fog node) and applies data fusion techniques to extract features. For our architecture they form the main component. Obviously a fog node has less power and a less global data view than a cloud node and hence it can apply less sophisticated data aggregation algorithms. It sends the transformed and fused data to the cloud for further processing and storage if required (ideally the fog node can make the ultimate decision). In an investigation into energy limitation, the authors in \cite{bormann2014terminology} found that these devices have restricted energy for particular tasks.
   
\item \textbf{Cloud node}s reside in the cloud and provide the final processing mechanism, obtaining the transformed data from fog nodes. They mainly apply machine learning algorithms and store the data. It is clear that the processing power and storage capability of the cloud is high. This power can be used even more effectively by using the presented approach. According to \cite{bormann2014terminology}, there is no energy limitation in the devices which are in cloud.
\end{itemize}

\subsection{Activity Recognition Using Accelerometer Traces}

To validate our architecture we have used the WISDM \cite{kwapisz2011activity} data set which is a set of accelerometer data on mobiles (particularly Android based) from 36 users who are doing 6 activities (walking, jogging, climbing upstairs, descending downstairs, sitting and standing). These users carried their mobiles while they were performing these activities for a fixed time. 

We divided the data into 10 seconds chunks. In addition, 43 features are created depending on 200 readings within the specified chunks. The transformed data contains 5418 accelerometer traces from the 36 users, with in average 150.50 traces per user and a standard deviation of 44.73. 

We conducted 3 sets of experiments: Firstly, we apply analytical algorithms on the transformed data in the cloud to calculate the accuracy of each algorithms and the execution time as a baseline. Secondly, we apply analytical algorithms on the transformed data in a fog gateway to calculate the execution time and to check the feasibility of the resource constraint devices while processing the data. Finally, we apply data aggregation algorithms on the raw data to extract features. The final approach minimises the data as much possible in the fog then sends the transformed data to the cloud for analysis. We measure the accuracy and execution time as well as the data amount send to the cloud.

Our hope was that a similar accuracy can be achieved with the third approach without increasing processing time and with significantly  reducing network data transmissions. 

  \section{Evaluation and Discussion}
 \subsection{Experimental Set up}
   As mentioned earlier it is not possible to apply classification algorithms on raw data which is time series data. Therefore, there is a need to transform raw data into features \cite{kwapisz2011activity}. In our experiment, we used a Raspberry Pi 3 model as an example of a low power fog gateway. The used Raspberry Pi has 1GB RAM and runs Raspbian Jessie with Pixel installed as operating system. In addition, to simulate the cloud device we used a 16GB RAM Linux System. We used the weka tool on both sides and we adjusted the heap size in both cloud and fog. In the fog the heap size was 650MB, in the cloud we allowed 8GB RAM for our experiment. Moreover, we used the same data aggregation methods that were used to extract features in \cite{kwapisz2011activity} to allow for comparability. We run data said aggregation methods in the Raspberry Pi to generate meaningful features depending on 200 readings where each has x,y and z acceleration information.
   
When we used the statistical measurements that are used in \cite{kwapisz2011activity} we created 43 features including the average of each axis, standard deviation of each axis, average absolute difference of each axis, average resultant acceleration for all axis, time between peaks of each axis and binned distribution for every axis (10 equal sized bins and totally 30 bins).

After the data is prepared we applied five classification methods from the Weka data mining and machine learning tools. The methods include decision tree (J48), logistic regression, multilayer perceptron, and naive Bayesian. Throughout our experiment we have used 10 fold cross validation.
 
 \subsection{Results}

   \begin{figure*}[htb]
  \centering
 % \vspace{-0.83cm}
  \includegraphics[scale=0.55]{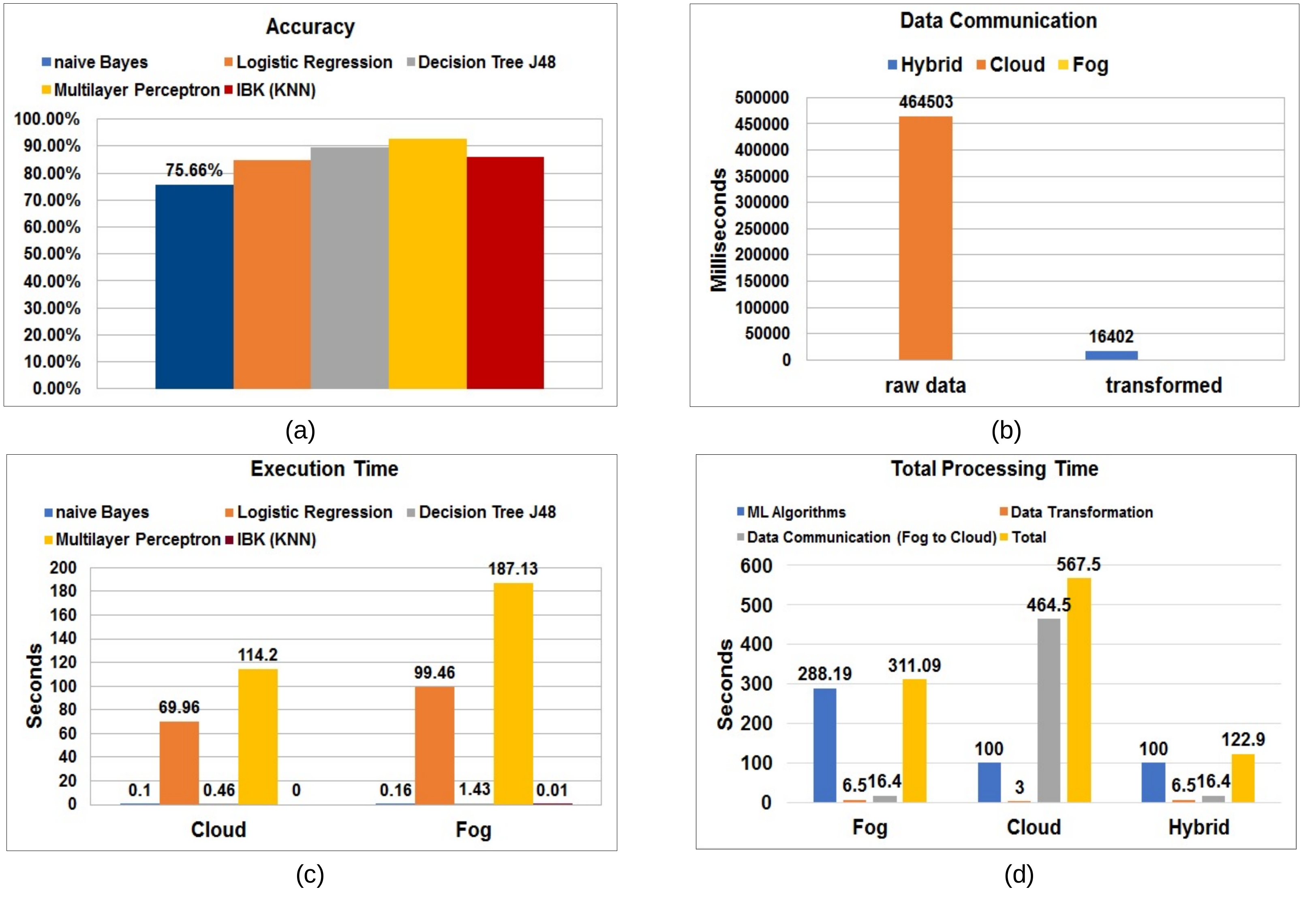}
 % \vspace{-0.21cm}	
  \caption{Data Processing Results (a - d)}
 \label{Figure:Scenario}	
 %\vspace{-0.23cm}	
 \end{figure*}
 
Figure \ref{Figure:Scenario}.$(a)$ shows the accuracy results of the 5 analysis algorithms that we applied on the transformed data. It is clear that from results that the multilayer perceptron has the highest accuracy percentage (100\% is the best result from the in Cloud analysis on raw data).

Figure \ref{Figure:Scenario}.$(b)$ shows the data communication time over network from fog (Raspberry Pi) to cloud. There are two bars visible: one for raw data and the other for transformed data. While applying this experiment the upload speed of the internet was 1Mbps. It is clear that the fog only device has no data communication cost because the processing happened in the device and no communication to the cloud took place. However, in the cloud approach the raw data communication over network from fog to cloud is extremely high, whereas in the hybrid approach the transformed data communication over the network from fog to cloud is low. This not surprising result confirms that we can save significantly on data communications by aggregating and pre-processing data early in the chain. 

Figure \ref{Figure:Scenario}.$(c)$ illustrates the execution time of the 5 analytics algorithms in both the cloud and fog device. The results show us that two algorithms (logistic regression and multilayer perceptron) have significant differences between the two sides. Obviously, the IoT device takes more time than the cloud to execute analytics algorithms because of its resource constraints. 

Figure \ref{Figure:Scenario}.$(d)$ demonstrates the total processing time for the three architectures. There are three measurements for each architecture including the execution time of analytics (ML) algorithms, the execution time of the data transformation process and the data communication time between local device and cloud. This graph needs a bit more explanation as the results are more interesting, so the details are as follows: 

\begin{itemize}
\item Fog (Raspberry PI):  The data transformation process is conducted locally and it is clear that the processing time is higher than in cloud. The analytics algorithms have been processed locally and they took much more time than cloud because of the processing power. However, data communication (the time to send data to the cloud) is very low as only aggregated data is being sent for storage. So, overall processing time is in the middle of the measured approaches.

\item Cloud: Data Communication is the time that the raw data takes from IoT device to cloud, which is clearly high as all raw data is being transmitted. The data transformation process was done in the cloud and due to the available resource runs quickly. Also, the analytics algorithms have been processed in the cloud and they took much less time than the fog because of the processing power. However, overall due to the significant amount of transmission time the cloud is the slowest approach in the given setting.

\item Hybrid: Here the data transformation process is done locally (in the fog) on the Raspberry Pi, with the usual observation. Data Communication is the time the transformed data takes from IoT device to cloud which as before is low. Finally the analytics algorithms have been processed in the cloud on the transformed data they took much less time than locally because of the processing power. Overall by combining the various strength this leads to a very good execution time.
\end{itemize}

Both fog and hybrid approaches looks similar to each other in most cases. However, they differ in terms of the place for applying the machine learning algorithms on the transformed data. In the fog approach the whole processing (data fusion and machine learning algorithms) happens in the node itself which can be considered as decentralised architecture. Whereas, in the hybrid approach the raw data is firstly fused in the fog node to extract features, and then these features are sent to the cloud for applying the machine learning algorithms. The major benefit of the hybrid approach over the fog one is using the power of the cloud for applying machine learning algorithms which need more processing power. Therefore, this step helps in reducing the processing time as a contributor to the overall data processing time.

These initial results show us that the proposed hybrid approach is good enough for the chosen dataset and analytical methods. It is clear from the results that data communication is efficient and provides significant gains. 
 
\textbf{Observation 1: Data consumption over network.} It is accepted that when the size of data is large the data consumption will be more expensive. The raw data was around 1 million rows which is equal to approximately 50 MB. However, after aggregating data into features by using data aggregation algorithms, the number of rows became 5418 rows and the size became 1.2 MB. This means that very significant savings to data transmission and storage can be made by early aggregation. This observation will gradually gain in importance as the number and quality of sensors increases rapidly and thus the rate and resolution at which data will be delivered grows quickly. By fusing the data locally before sending it to the cloud, we are not only reducing the data we are also determining which data is meaningful and only send that. This will reduce the energy consumption of fog and sensor devices which typically gain internet connectivity through 3, 4 or 5G, thus batteries in the devices will last longer.

\textbf{Observation 2: Accuracy.} Aggregated data leads to less accuracy in the results compared to working with raw data. All presented approaches are effected in the same way. Overall, the loss of accuracy is not drastic: the lowest bar is 75\% with the highest being around 93\%. Clearly the used analysis method has an impact with trade-offs such as the used local processing power as well as methods optimised for this localised setting being factors that can influence the accuracy. The right balance will in terms of privacy, accuracy, resource cost, energy consumption and data transmission will need to be identified and our future work will further this area.

Out of these experiments and observations we can conclude that one important aspect for future work is a development that combines the ideas of distributed data aggregation with analysis methods that can also be distributed effectively and be run in low power environments. The fact that they can operate on smaller data sets will help, but somehow the methods need to contain a core part based on global understanding.  

Aside. For completeness we like to note that we initially attempted to conduct the experiments with a Raspberry Pi model b with 512MB RAM, however this was not capable to apply some of the weka toolbox analysis algorithms because of the RAM constraint. Hence we used a slightly more powerful version as reported above. 
 
\section{Related work}
In recent years, the main cloud providers have been promising new IoT services with various functionalities and advantages. One of the main cloud providers is Microsoft with its Azure Stack \cite{azure} which offers a hybrid cloud that allows companies to transfer benefits from their servers while keeping the management of servers for new types of cloud (hybrid cloud). In addition, they provide gateway devices in the cloud and data analytics. Similarly, IBM has an online web analytics system with IBM Digital Analytics. This service provides tracking and analysing of behaviours from visitors. The data analytics uses high power servers inside IBM. The IBM PureData system promises fast data analytics and warehouse that combine warehouse, data centres and analytics \cite{coyne2017ibm}. Although the above systems are promising powerful data analytics approaches, they do not support  a fog gateway concept which resides between IoT devices and the cloud. As we can see from our results that uploading high volumes of raw data consumes time and energy. Therefore, a fog gateway concept is important for real time services to save time, energy and resource cost.

Data fusion is an active area in research and business particularly with a view to optimised data analytics. There are several data fusion techniques that focus on reducing the consumption of energy in \cite{dhasian2013survey,abdelgawad2012resource}. They have used a variety of methods including fuzzy set theory and neural networks. They succeeded in terms of removing redundancies while fusing the data. However, they did not focus on the resource constraints of devices that embed the sensors. In contrast, they assume that these devices work efficiently without a need to pay attention to their limitations. More importantly, these mechanisms send all the data to centralised computation systems, which affects the data communication cost, privacy and energy as well. As we could see in our experiments sending the raw data to the cloud is not efficient in terms of data communication over the network.

\section{Conclusions and Future Work}

This paper presents a hybrid approach in which data is fused in the fog before being send to the cloud to reduce data communication over the network. The results show  that this architecture is successful in terms of reducing data communication cost over network without significantly reducing accuracy of later decision making. We presented the proposed approach and its relevant methods. In addition, we used the WISDM dataset \cite{kwapisz2011activity} to validate our architecture. 

On the basis of the promising findings presented in this paper, future work will involve creating different features with different algorithms for better data aggregation further reducing data communication while attempting to increase (or at least maintain) accuracy. One particular key piece of work has been alluded to in the results section: the development of analysis methods that can be distributed and executed in efficient ways on low power devices. A key strategy here will be the exploration of the balance of analysis with limited (local) data sets vs the availability of a global view. In addition, evaluation of energy consumption and consideration of the positive impact on privacy will be aspects of future work. Furthermore, additional datasets to test and evaluate our hybrid approach further will investigated.

\section{Acknowledgments}
Dr. Charith Perera's work is funded by EPSRC award number DERC EP/M023001/1 (Digital Economy Research Centre). Badraddin Alturki's research is funded by Saudi Arabian Cultural bureau in London and his scholarship is granted by King Abdul Aziz University.

% Balancing columns in a ref list is a bit of a pain because you
% either use a hack like flushend or balance, or manually insert
% a column break.  http://www.tex.ac.uk/cgi-bin/texfaq2html?label=balance
% multicols doesn't work because we're already in two-column mode,
% and flushend isn't awesome, so I choose balance.  See this
% for more info: http://cs.brown.edu/system/software/latex/doc/balance.pdf
%
% Note that in a perfect world balance wants to be in the first
% column of the last page.
%
% If balance doesn't work for you, you can remove that and
% hard-code a column break into the bbl file right before you
% submit:
%
% http://stackoverflow.com/questions/2149854/how-to-manually-equalize-columns-
% in-an-ieee-paper-if-using-bibtex
%
% Or, just remove \balance and give up on balancing the last page.
%
\balance{}

% BALANCE COLUMNS
\balance{}

% REFERENCES FORMAT
% References must be the same font size as other body text.
\bibliographystyle{SIGCHI-Reference-Format}
\bibliography{library}


\begin{thebibliography}{00}

%%% the Web address.

\ifx \showCODEN    \undefined \def \showCODEN     #1{\unskip}     \fi
\ifx \showDOI      \undefined \def \showDOI       #1{{\tt DOI:}\penalty0{#1}\ }
  \fi
\ifx \showISBNx    \undefined \def \showISBNx     #1{\unskip}     \fi
\ifx \showISBNxiii \undefined \def \showISBNxiii  #1{\unskip}     \fi
\ifx \showISSN     \undefined \def \showISSN      #1{\unskip}     \fi
\ifx \showLCCN     \undefined \def \showLCCN      #1{\unskip}     \fi
\ifx \shownote     \undefined \def \shownote      #1{#1}          \fi
\ifx \showarticletitle \undefined \def \showarticletitle #1{#1}   \fi
\ifx \showURL      \undefined \def \showURL       #1{#1}          \fi

\bibitem{abdelgawad2012data}
{Ahmed Abdelgawad} {and} {Magdy Bayoumi}. 2012a.
\newblock \showarticletitle{Data fusion in WSN}.
\newblock In {\em Resource-aware data fusion algorithms for wireless sensor
  networks}. Springer, 17--35.
\newblock


\bibitem{abdelgawad2012resource}
{Ahmed Abdelgawad} {and} {Magdy Bayoumi}. 2012b.
\newblock {\em Resource-Aware data fusion algorithms for wireless sensor
  networks}. Vol. 118.
\newblock Springer Science \& Business Media.
\newblock


\bibitem{abu2013data}
{Mervat Abu-Elkheir}, {Mohammad Hayajneh}, {and} {Najah~Abu Ali}. 2013.
\newblock \showarticletitle{Data management for the internet of things: Design
  primitives and solution}.
\newblock {\em Sensors\/} {13}, 11 (2013), 15582--15612.
\newblock


\bibitem{andrade_gedik_turaga_2014}
{Henrique C.~M. Andrade}, {Buğra Gedik}, {and} {Deepak~S. Turaga}. 2014.
\newblock {\em Fundamentals of Stream Processing: Application Design, Systems,
  and Analytics}.
\newblock Cambridge University Press.
\newblock
\showDOI{%
\url{http://dx.doi.org/10.1017/CBO9781139058940}}


\bibitem{bormann2014terminology}
{Carsten Bormann}, {Mehmet Ersue}, {and} {A Keranen}. 2014.
\newblock {\em Terminology for constrained-node networks}.
\newblock {T}echnical {R}eport.
\newblock


\bibitem{castanedo2013review}
{Federico Castanedo}. 2013.
\newblock \showarticletitle{A review of data fusion techniques}.
\newblock {\em The Scientific World Journal\/}  {2013} (2013).
\newblock


\bibitem{chhabra2015data}
{Sakshi Chhabra} {and} {Dinesh Singh}. 2015.
\newblock \showarticletitle{Data Fusion and Data Aggregation/Summarization
  Techniques in WSNs: A Review}.
\newblock {\em International Journal of Computer Applications\/} {121}, 19
  (2015).
\newblock


\bibitem{coyne2017ibm}
{Larry Coyne}, {Joe Dain}, {Phil Gilmer}, {Patrizia Guaitani}, {Ian Hancock},
  {Antoine Maille}, {Tony Pearson}, {Brian Sherman}, {Christopher Vollmar},
  {and} {others}. 2017.
\newblock {\em Ibm private, public, and hybrid cloud storage solutions}.
\newblock IBM Redbooks.
\newblock


\bibitem{dasarathy1997sensor}
{Belur~V Dasarathy}. 1997.
\newblock \showarticletitle{Sensor fusion potential exploitation-innovative
  architectures and illustrative applications}.
\newblock {\it Proc. IEEE} {85}, 1 (1997), 24--38.
\newblock


\bibitem{dhasian2013survey}
{Hevin~Rajesh Dhasian} {and} {Paramasivan Balasubramanian}. 2013.
\newblock \showarticletitle{Survey of data aggregation techniques using soft
  computing in wireless sensor networks}.
\newblock {\em IET Information Security\/} {7}, 4 (2013), 336--342.
\newblock


\bibitem{hall1997introduction}
{David~L Hall} {and} {James Llinas}. 1997.
\newblock \showarticletitle{An introduction to multisensor data fusion}.
\newblock {\it Proc. IEEE} {85}, 1 (1997), 6--23.
\newblock


\bibitem{harrington2012machine}
{Peter Harrington}. 2012.
\newblock {\em Machine learning in action}. Vol.~5.
\newblock Manning Greenwich, CT.
\newblock


\bibitem{kwapisz2011activity}
{Jennifer~R Kwapisz}, {Gary~M Weiss}, {and} {Samuel~A Moore}. 2011.
\newblock \showarticletitle{Activity recognition using cell phone
  accelerometers}.
\newblock {\em ACM SigKDD Explorations Newsletter\/} {12}, 2 (2011), 74--82.
\newblock


\bibitem{ng2014engineering}
{IC Ng}. 2014.
\newblock \showarticletitle{Engineering a Market for Personal Data: The
  Hub-of-all-Things (HAT), A Briefing Paper}.
\newblock {\em WMG Service Systems Research Group Working Paper Series\/}
  (2014).
\newblock


\bibitem{7004800}
{C. Perera}, {C.~H. Liu}, {and} {S. Jayawardena}. 2015.
\newblock \showarticletitle{The Emerging Internet of Things Marketplace From an
  Industrial Perspective: A Survey}.
\newblock {\em IEEE Transactions on Emerging Topics in Computing\/} {3}, 4 (Dec
  2015), 585--598.
\newblock
\showISSN{2168-6750}
\showDOI{%
\url{http://dx.doi.org/10.1109/TETC.2015.2390034}}


\bibitem{preden2015data}
{Jurgo Preden}, {Jaanus Kaugerand}, {Erki Suurjaak}, {Sergei Astapov}, {Leo
  Motus}, {and} {Raido Pahtma}. 2015.
\newblock \showarticletitle{Data to decision: pushing situational information
  needs to the edge of the network}. In {\em Cognitive Methods in Situation
  Awareness and Decision Support (CogSIMA), 2015 IEEE International
  Inter-Disciplinary Conference on}. IEEE, 158--164.
\newblock


\bibitem{sundmaeker2010vision}
{Harald Sundmaeker}, {Patrick Guillemin}, {Peter Friess}, {and} {Sylvie
  Woelffl{\'e}}. 2010.
\newblock \showarticletitle{Vision and challenges for realising the Internet of
  Things}.
\newblock {\em Cluster of European Research Projects on the Internet of Things,
  European Commision\/} (2010).
\newblock


\bibitem{Vermesan2009}
{Ovidiu Vermesan}, {Peter Friess}, {Patrick Guillemin}, {Sergio Gusmeroli},
  {Harald Sundmaeker}, {Alessandro Bassi}, {Ignacio~Soler Jubert}, {Margaretha
  Mazura}, {Mark Harrison}, {Markus Eisenhauer}, {Pat Doody}, {Friess Peter},
  {Guillemin Patrick}, {Gusmeroli Sergio}, {Bassi {Harald, Sundmaeker
  Alessandro}}, {Jubert {Ignacio Soler}}, {Mazura Margaretha}, {Harrison Mark},
  {Eisenhauer Markus}, {and} {Doody Pat}. 2009.
\newblock \showarticletitle{{Internet of Things Strategic Research Roadmap}}.
\newblock {\em Internet of Things Strategic Research Roadmap\/} (2009), 9--52.
\newblock
\showISBNx{978-87-92329-67-7}
\showISSN{0036-8733}
\showDOI{%
\url{http://dx.doi.org/pdf/IoT_Cluster_Strategic_Research_Agenda_2011.pdf}}


\bibitem{wang2015city}
{Meisong Wang}, {Charith Perera}, {Prem~Prakash Jayaraman}, {Miranda Zhang},
  {Peter Strazdins}, {and} {Rajiv Ranjan}. 2015.
\newblock \showarticletitle{City data fusion: Sensor data fusion in the
  internet of things}.
\newblock {\em arXiv preprint arXiv:1506.09118\/} (2015).
\newblock


\bibitem{white1991data}
{Franklin~E White}. 1991.
\newblock {\em Data fusion lexicon}.
\newblock {T}echnical {R}eport. DTIC Document.
\newblock


\bibitem{azure}
{J. Woolsey}. 2016.
\newblock \showarticletitle{Powering the Next Generation Cloud with Azure
  Stack}.
\newblock  (2016).
\newblock


\bibitem{xhafa2014semantics}
{Fatos Xhafa} {and} {Leonard Barolli}. 2014.
\newblock Semantics, intelligent processing and services for big data.
\newblock   (2014).
\newblock


\bibitem{yan2015trustworthy}
{Zheng Yan}, {Jun Liu}, {Athanasios~V Vasilakos}, {and} {Laurence~T Yang}.
  2015.
\newblock \showarticletitle{Trustworthy data fusion and mining in Internet of
  Things}.
\newblock {\em Future Generation Computer Systems\/} {49}, C (2015), 45--46.
\newblock


\bibitem{zaslavsky2013sensing}
{Arkady Zaslavsky}, {Charith Perera}, {and} {Dimitrios Georgakopoulos}. 2013a.
\newblock \showarticletitle{Sensing as a service and big data}.
\newblock {\em arXiv preprint arXiv:1301.0159\/} (2013).
\newblock


\bibitem{DBLP:journals/corr/abs-1301-0159}
{Arkady~B. Zaslavsky}, {Charith Perera}, {and} {Dimitrios Georgakopoulos}.
  2013b.
\newblock \showarticletitle{Sensing as a Service and Big Data}.
\newblock {\em CoRR\/}  {abs/1301.0159} (2013).
\newblock
\showURL{%
\url{http://arxiv.org/abs/1301.0159}}


\end{thebibliography}

\end{document}